 \documentclass[12pt]{article}    
 \usepackage{graphicx,colordvi,epsf,rotate}
 \usepackage{graphicx}
 \usepackage{epsfig}

\begin{document}
  \parskip 6pt

\begin{center} {\sf For Electronic Journal of Theoretical Physics,
  Landau Issue 2008. ~05/06/2008}  \\[5mm]
{\Large\bf The Vortex Lattice in Ginzburg-Landau
    Superconductors}\\[4mm]

           Ernst Helmut Brandt\\[2mm]
 Max Planck Institute for Metals Research, D-70506 Stuttgart, Germany
\end{center}

\begin{abstract}
 Abrikosov's solution of the linearized Ginzburg-Landau theory
describing a periodic lattice of vortex lines in type-II
superconductors at large inductions, is generalized to
non-periodic vortex arrangements, e.g., to lattices with a
vacancy surrounded by relaxing vortices and to periodically
distorted lattices that are needed in the nonlocal theory of
elasticity of the vortex lattice. Generalizations to lower
magnetic inductions and to three-dimensional arrangements
of curved vortex lines are also given. Finally, it is shown
how the periodic vortex lattice can be computed for bulk
superconductors and for thick and thin films in a perpendicular
field for all inductions $\bar B$ and Ginzburg-Landau
parameters $\kappa$.
\end{abstract}

\section{Introduction}  

  From Ginzburg-Landau (GL) theory \cite{1} Landau's thesis student
Alexei Abrikosov predicted that superconductors with  a GL
parameter $\kappa > 1/\sqrt2$ may contain a lattice of vortices of
supercurrent, or flux lines (fluxons) with quantized magnetic flux
$\Phi_0 = h/2e = 2\cdot 10^{-15}$ Tm$^2$. Abrikosov had linearized
the GL equations with respect to a small order parameter
$|\psi|^2$ and discovered a solution $\psi(x,y)$ possessing
a regular lattice of zero lines.
This lattice solution appears when the applied magnetic field
$B_a$ (along $z$) is decreased below the upper critical field
$B_{c2} = \Phi_0 / (2\pi \xi^2)$, where $\xi = \lambda/\kappa$
is the GL coherence length. At $B_a = B_{c2}$ one has the
average induction $\bar B = B_{c2}$. With decreasing $B_a$,
the induction decreases and reaches $\bar B = 0$ at the
lower critical field $B_a = B_{c1} = \Phi_0 ( \ln \kappa +
 \alpha) /(4\pi \lambda^2)$  with $\alpha(\kappa) \approx 0.5$
for $\kappa \gg 1$ (see below). At the same time the vortex
lattice spacing $a \approx (\Phi_0/\bar B)^{1/2}$ increases and
diverges when $\bar B \to 0$.
Abrikosov tells that he had obtained this vortex solution in
1953 but Landau didn't like it, stating that there are no
line-like singularities in electrodynamics. Only when
Feynman \cite{2} had published his paper on vortices in
superfluid Helium, did Landau agree and Abrikosov could publish
his solution in 1957 \cite{3}. For his prediction of the
vortex lattice Abrikosov 50 years later in 2003 received the
Nobel Prize in Physics together with Vitalii Ginzburg and
Anthony Leggett.

  After first evidence of the triangular vortex lattice in
superconducting Niobium by small-angle neutron scattering in
Saclay \cite{4}, Tr\"auble and Essmann in Stuttgart
succeeded \cite{5,6} to observe the vortex lattice directly
by decorating the surface of a superconductor with iron
microcrystallites (``magnetic smoke''). At that time I joined
this research group headed by A.~Seeger and wrote my thesis on
the theory of defects in the vortex lattice \cite{7}.
Parts 1 and 2 deal with low inductions $\bar B \ll B_{c2}$
when London theory may be used and the vortices interact
with each other pairwise, similar to 2D atomic lattices.
Parts 3 and 4 consider high inductions $\bar B \approx B_{2}$,
where the shape of the GL solutions $\psi(x,y)$ and $B(x,y)$
may be obtained from linearized GL theory, while the nonlinear
GL terms determine the amplitudes of this $\psi$ and $B$.
My thesis extended Abrikosov's theory of periodic vortex lattices
to non-periodic vortex arrangements, see below.
  Such distorted-lattice solutions are required to calculate the
elastic energy of the vortex lattice and the energy of lattice
defects like vacancies and dislocations. They are also helpful
to visualize where the solutions of the linearized GL theory
apply and how they have to be modified at lower inductions.

\section{Abrikosov's ideal vortex lattice near $B_{c2}$}  

  In the usual reduced units (length $\lambda$, induction
$\sqrt2 B_c$, energy density $B_c^2/\mu_0$, where
$B_c = B_{c2}/\sqrt2 \kappa$ is the thermodynamic critical field)
the spatially averaged free
energy density $F$ of the GL theory referred to the Meissner
state ($\psi=1$, ${\bf B} =0$) within the superconductor reads
\begin{equation}  
  F = \left\langle {(1-|\psi|^2)^2 \over 2} + \left| \left(
  {\nabla \over i\kappa} - {\bf A} \right) \psi \right|^2
  + {\bf B}^2 \right\rangle .
  \end{equation}
Here $\psi({\bf r}) = f\, \exp(i\varphi)$ is the complex
GL function, ${\bf B(r)}=\nabla\times {\bf A}$ the magnetic
induction, ${\bf A(r)}$ the vector potential, and
$\langle \dots \rangle = (1/V) \int_V d^3r \dots $ means
spatial averaging over the superconductor with volume $V$.
Introducing the super velocity
${\bf Q}({\bf r}) = {\bf A} -\nabla\varphi/\kappa$ and the
magnitude $f({\bf r}) = |\psi|$ one may write $F$ as a
functional of the real and gauge-invariant
functions $f$ or $f^2 = \omega$ and ${\bf Q}$,
  \begin{equation}  
  F = \left\langle {(1-f^2)^2 \over 2} + {(\nabla f)^2 \over
  \kappa^2 } +f^2 Q^2 +(\nabla\! \times\! {\bf Q})^2 \right\rangle.
  \end{equation}
In the presence of vortices ${\bf Q}({\bf r})$ has to be chosen
such that $\nabla\! \times\! {\bf Q}$ has the appropriate
singularities along the vortex cores, where $f$ vanishes.
By minimizing this $F$ with respect to $\psi$, ${\bf A}$ or
$f$, ${\bf Q}$, one obtains the GL equations together with the
appropriate boundary conditions.
For the superconducting film considered in Sec.~5, one has to add
the energy of the magnetic stray field outside the film, which
makes the perpendicular component $B_z$ of ${\bf B}$ continuous
at the film surface, see below.

   The two GL equations are obtained by minimization of $F$ (1)
with respect to $\psi$ and ${\bf A}$, $\delta F /\delta\psi =0$
and $\delta F / \delta{\bf A} = 0$, yielding
  \begin{eqnarray}  
   (\nabla/i -\kappa{\bf A})^2 \psi = \kappa (1-|\psi|^2)\psi\,,\\
   \nabla\times [\nabla\times {\bf A}] = |\psi|^2 {\bf Q} \,.
  \end{eqnarray}
With $B_a$ and $\bar B$ chosen along the $z$ axis and in the
gauge $A_x = -\bar B y +\tilde A_x(x,y)$, $A_y = \tilde A_y(x,y)$
($\tilde A_x$, $\tilde A_y$ are terms of higher order)
the linearized first GL equation, obtained by omitting the term
$|\psi|^2$ in (3), has the general solution
  \begin{eqnarray}  
  \psi(x,y) = \exp(-\kappa \bar B y^2 /2)~ g(x,y) \,,\\
  {\partial g \over\partial x} +i{\partial g \over\partial y}=0 \,.
  \end{eqnarray}
This means $g(x,y) = g(z)$, $z=x+iy$,  can be any analytical
function. For a periodic solution satisfying
$|\psi|^2({\bf r +R}_{mn}) =|\psi|^2({\bf r})$, ${\bf r} =(x,y)$,
with real and reciprocal lattice vectors
  \begin{eqnarray}  
    {\bf R}_{mn} =(mx_1 +nx_2;\, ny_2) \,,\\
    {\bf K}_{mn} =(2\pi/x_1y_2)(my_2;\, -mx_2 +nx_1) \,,
  \end{eqnarray}
($m,n = 0, \pm1, \pm2, \dots$; triangular lattice: $x_1=a$,
$x_2=x_1/2$, $y_2=x_1\sqrt3/2$; square lattice: $x_1=y_2=a$,
$x_2=0$) and with a zero at ${\bf r}=0$, one obtains for $g(z)$
the function $\vartheta_1$ defined as \cite{8}
  \begin{equation}  
   \vartheta_1(z,\tau) = 2\sum_{n=0}^\infty (-)^n \exp[\,i\pi \tau
   (n + {\textstyle{ 1\over 2}}  )^2\,] \sin(2n+1)z \,.
  \end{equation}
Thus, the periodic Abrikosov solution with zeros at the
${\bf r=R}_{mn}$ may be written as
  \begin{eqnarray}  
  \psi_A(x,y) = \exp\Big(\! -{\pi y^2 \over x_1y_2} \Big)~\vartheta_1
  \Big(\, {\pi\over x_1} (x+iy),\, {x_2+iy_2 \over x_1}\, \Big) \,.
  \end{eqnarray}
This solution has the mean induction $\bar B = \Phi_0/(x_1y_2)$,
normalized order parameter $\langle |\psi_A|^2 \rangle =1$, and the
Fourier series $|\psi_A|^2 = \omega_A(x,y)$,
  \begin{eqnarray}  
  \omega_A({\bf r}) = \sum_{{\bf K}_{mn}} (-)^{mn+m+n} \exp
  \Big( -{ K_{mn}^2 x_1y_2 \over 8\pi} \Big)\,  e^{i{\bf K r}} \,.
  \end{eqnarray}
From the zero $\omega_A(0,0) = 0$ follows that the sum over all
Fourier coefficients in (11) is zero for all lattice symmetries.
(Abrikosov \cite{3} chose a different position for $\omega_A =0$
and thus obtained the function $\vartheta_3$ \cite{8}).

\section{Distorted vortex lattice near $B_{c2}$}  

  The GL solution for a distorted vortex lattice near $B_{c2}$ is
obtained as follows. Assume that each of the straight and parallel
vortex lines is displaced from its ideal lattice positions
${\bf R}_{mn} = {\bf R}_\nu =(X_\nu, Y_\nu)$
by displacements ${\bf s}_\nu = (s_{\nu x}, s_{\nu y})$,
${\bf r}_\nu = (x_\nu, y_\nu) = {\bf R}_\nu +{\bf s}_\nu$,
such that the displacement field itself is periodic with
a super lattice $N$ times larger than the vortex lattice, but
with same symmetry, ${\bf s(r} +N{\bf R}_{mn}) = {\bf s(r)}$.
Where needed we use a continuous displacement field ${\bf s(r)}$
defined such that it has the same Fourier transform as the
discrete ${\bf s}_\nu = {\bf s(R}_\nu)$.  A distorted triangular
vortex lattice with spacing $x_1 = a$ then has the solution, Eq.~(5),
  \begin{equation}  
  \psi(x,y) = c_1 \exp\Big(\! -{2 \pi y^2 \over \sqrt3 a^2} \Big)\,
  \vartheta_1({\pi \over a} z, \tau) \prod_\nu {
  \vartheta_1[ (\pi / Na) (z-z_\nu -s_\nu), \tau ] \over
  \vartheta_1[ (\pi / Na) (z-z_\nu),        \tau ] }
  \end{equation}
with $z=x+iy$, $z_\nu = X_\nu + iY_\nu$, $\tau = (1 +i \sqrt3)/2$,
$s_\nu = s_{x\nu} +i s_{y\nu}$, the product is over one super cell,
and $c_1 \approx 1$ is a
normalization constant. When all $s_\nu = 0$, the product in (12)
is unity, $\prod_\nu =1$, thus the first two factors in (12)
are the ideal lattice solution with $c_1=1$, cf.~Eq.~(10). Each
factor of the product shifts one
zero from ${\bf r=R}_\nu $ to ${\bf r=R}_\nu +{\bf s}_\nu$.
The absolute value of $|\psi|^2 = \omega$ of this GL function may
also be expressed in terms of the Fourier series
$\omega_A({\bf r})$ (11),
  \begin{equation}  
  \omega({\bf r}) = c_1^2 \,\omega_A({\bf r}) \prod_\nu {
  \omega_A[ ({\bf r-R}_\nu -{\bf s}_\nu) /N ]    \over
  \omega_A[ ({\bf r-R}_\nu             ) /N ] } \,.
  \end{equation}
In the limit of infinite super cell, $N \to \infty$, one may use
$\vartheta_1(z/N, \tau) \propto z/N$ for $|z|/N \ll 1$,
thus one may replace the function $\vartheta_1$ by its argument
since all the constant factors cancel or combine to a normalization
factor that follows from numerics. One then obtains simply
  \begin{equation}  
  \omega({\bf r}) = c_1^2~ \omega_A({\bf r}) \prod_\nu {
  | {\bf r-R}_\nu -{\bf s}_\nu |^2 \over | {\bf r-R}_\nu |^2 } \,.
  \end{equation}

\begin{figure}[tbh]  
\epsfxsize= .7\hsize  \vskip 1.0\baselineskip \centerline{
\epsffile{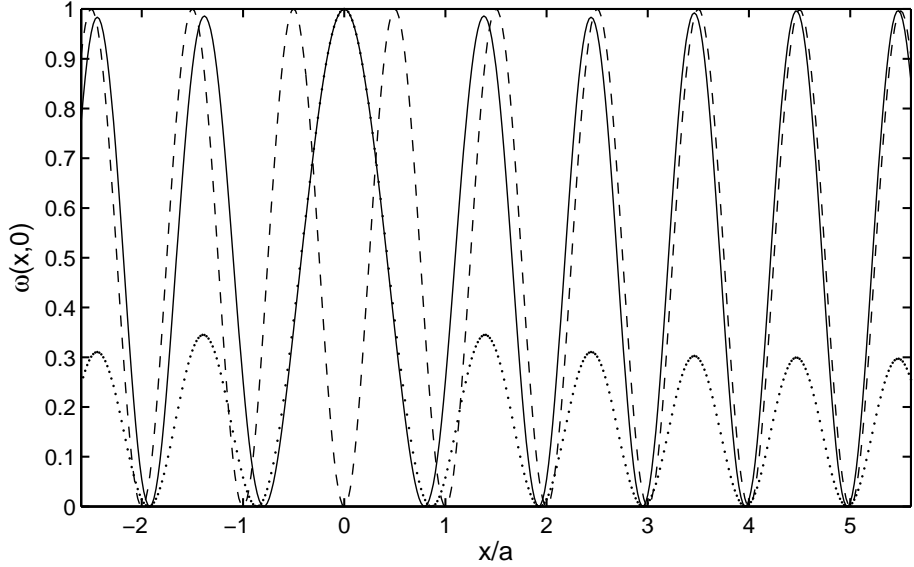}}  \vspace{ 0 mm}
\caption{\label{fig1} Order parameter $\omega(x,0)$ for the
ideal vortex lattice (dashed line) and for vortex lattice with
vacancy, Eq.~(15), with simple relaxation field (19) (dotted line)
and with better relaxation field that minimizes the defect energy
(solid line, see text).
 } \end{figure}   

\begin{figure}[tbh]  
\epsfxsize= .7\hsize  \vskip 1.0\baselineskip \centerline{
\epsffile{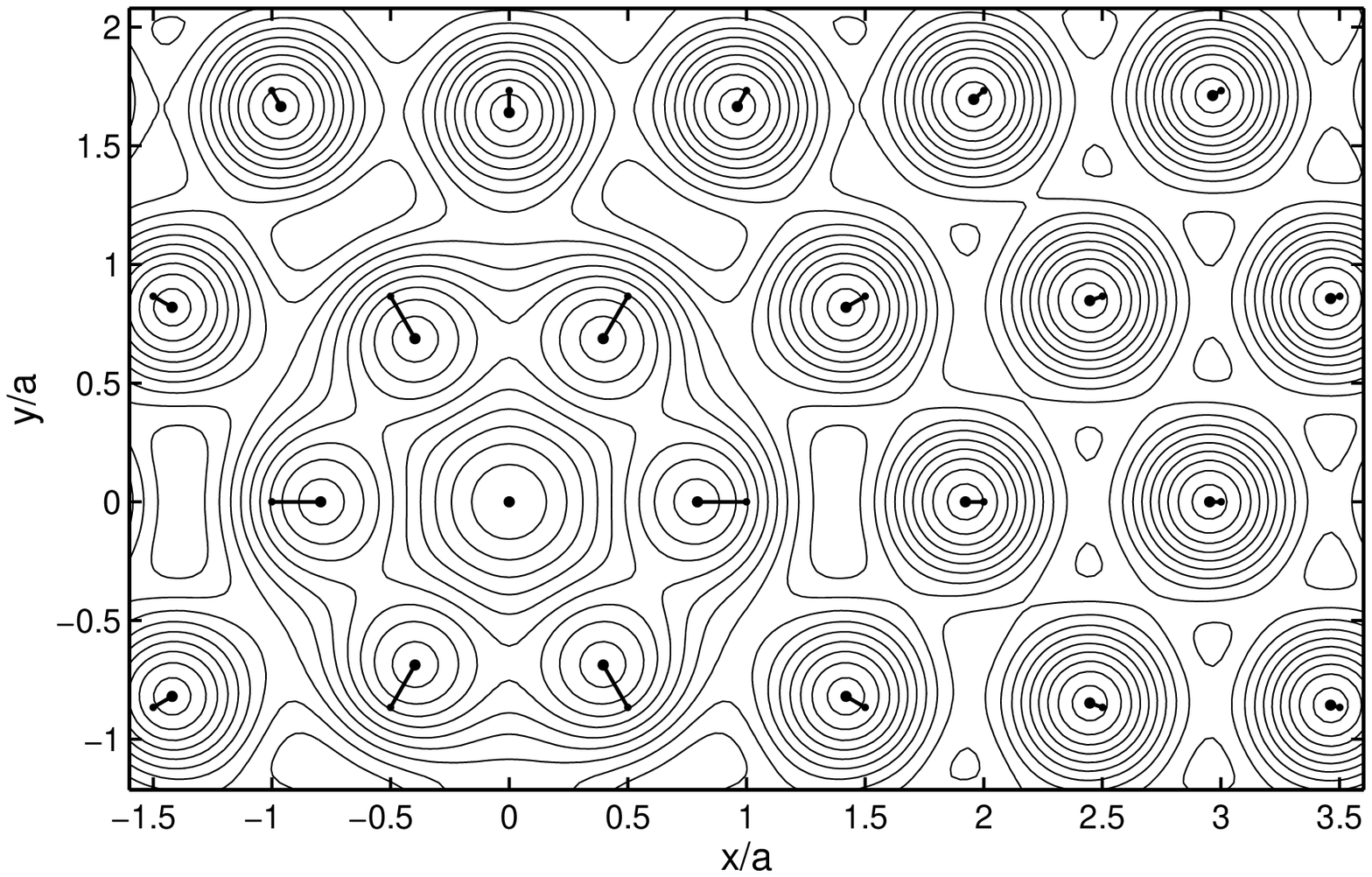}}  \vspace{ 0 mm}  
\caption{\label{fig2} Contour lines of the order parameter
$\omega(x,y)$ (15) of a vortex lattice with one vacancy at $x=y=0$
and complete relaxation, see solid line in Fig.~1. The vortex
displacements are indicated by short bold lines between two dots.
 } \end{figure}   

\section{Vortex lattice vacancy near $B_{c2}$}  

  Removing the central vortex at ${\bf R}_\nu=0$ adds a factor
$1/r^2$ to the linearized solution $\omega({\bf r})$. Obviously,
if the other vortices are not allowed to relax, this solution
at large distances vanishes as $1/r^2$; it cannot be normalized
and its energy is infinite. However, if the relaxation of the
other vortices is chosen appropriately it will minimize the defect
energy and make it finite. This can be seen from the solution
  \begin{eqnarray}  
  \omega({\bf r}) = c_1^2 \,{ \omega_A({\bf r}) \over r^2} \,
  \Big| {h(z) \over h(0)} \Big|^2  \,,~~
  h(z) = \prod_{\nu\ne 0} \Big( 1-{ s_\nu \over z-z_\nu } \Big) \,.
  \end{eqnarray}
The constant factor $|1/h(0)|^2$ was added to force convergence of
the infinite product. The solution for a super lattice of
vacancies positioned at the $N{\bf R}_\nu$, is given by
expression (13) divided by $\omega({\bf r}/N)$ that removes the
zeros at positions $N {\bf R}_\nu$..

   The energy of both the ideally periodic and the distorted
vortex lattices is calculated via the Abrikosov parameter
  \begin{eqnarray}  
  \beta = { \langle |\psi|^4 \rangle \over
            \langle |\psi|^2 \rangle^2 } =
 {\langle \omega^2 \rangle \over \langle\omega\rangle} \ge 1 \,.
  \end{eqnarray}
This $\beta$ enters the free energy of the linearized GL theory
(referred to the normal state), that has to be minimized when
$\bar B$ is held constant,
  \begin{equation}  
   F = {\bar B^2 \over 2\mu_0} -{(B_{c2}-\bar B)^2 \over
      2\mu_0 [1 + (2\kappa^2 - 1)\beta \, ]}
  \end{equation}
and the free enthalpy that has to be minimized when $B_a$ is
held constant,
  \begin{equation}  
   G = F - {\bar B B_a \over \mu_0} = -{(B_{c2}-B_a)^2 \over
       2\mu_0 (2\kappa^2 - 1)\beta }  \,.
  \end{equation}
The elastic energy of the distorted vortex lattice is the
product of the derivatives $\partial F / \partial \beta$ or
$\partial G / \partial \beta$ times the change of
$\beta\{ \psi \}$ times the volume, with the limit of infinite
volume taken. This means that all elastic energies and energies
of structural defects near $B_{c2}$ vanish as
$(B_{c2} -\bar B)^2 \propto (B_{c2} -B_a)^2$.
This is true also for the shear modulus
$c_{66}$ of the vortex lattice, which can be obtained using
Abrikosov's periodic lattice solution \cite{7,9}.

  In the case of the vortex vacancy, the resulting
defect energy is finite only if the vortices relax (shift
towards the removed vortex) such that at large distances (and
after numerical minimization practically at all distances)
the vortex displacements are
  \begin{equation}  
  {\bf s}_\nu = - {{\bf R}_\nu \over 2\pi n R_\nu^2 }
  \end{equation}
with $n = \bar B /\Phi_0 = 1/(x_1 y_2)$ the vortex density.
If the radial displacements were chosen smaller (larger) than
in (19), the order parameter (15) would vanish (diverge)
at large distances $r$. But with the correct displacements
that minimize $\beta$ and thus the defect energy, the
amplitude of the oscillating order parameter stays almost
constant, even near the vacancy. This can be seen in Fig.~1,
where the profiles of $\omega(x,y)$ along $y=0$ are plotted
for ideal triangular vortex lattice and for the lattice with
a central vacancy with the simple relaxation (19) and with
an improved relaxation field ${\bf s}_\nu = - {\bf R}_\nu
 [\sqrt{3} a^2 / (4\pi R_\nu^2) + 0.068 a^4/ R_\nu^4 ]$.
Figure 2 shows the contour lines of the fully relaxed
order parameter $\omega(x,y)$, which has a maximum at
the origin (vacancy position) and minima (zeros) at the
vortex positions.

 The displacement (19) means that the area of the ``relaxing
ring'', $2\pi R_\nu s_\nu = 1/n = x_1 y_2 = \Phi_0/\bar B$,
exactly equals the area of one lattice cell. In other words, the
removal of one vortex at ${\bf R}_\nu =0$ is compensated by the
relaxation of the surrounding vortices such that the average
vortex density in any contour (containing or not containing the
vacancy) stays constant and equals the density $n$ that was there
before the vacancy was introduced. Note that the field (19)
satisfies $\nabla\cdot {\bf s(r)} = 0$, and thus describes a pure
shear deformation. More precisely, one has $\nabla\cdot {\bf s(r)}
= (1/n) \delta_2({\bf r})$ ($\delta_2$ is the 2D delta function),
i.e., the displacement field (19) ``remembers'' that one vortex
cell area was removed.
Further interesting properties of structural defects in vortex
lattices and other two- or three-dimensional soft lattices are
discussed in [10].

\section{Distorted vortex lattice away from $B_{c2}$}  

  As shown with the vacancy example, the distorted-lattice
solution (14) of the linearized GL equations yields finite
energies of lattice defects only if most of the vortex
displacements are allowed to relax appropriately. But without
this relaxation, the elastic energy is infinite. For example,
if only the one vortex at the origin is displaced,
${\bf s}_\nu = s_0 \delta_{\nu 0} {\bf \hat x}$
($\delta_{\nu 0} =1$ if $\nu=0$, else $\delta_{\nu 0} =0$)
one has from (14)
  \begin{eqnarray}  
  \omega({\bf r}) = \omega_A({\bf r}) |{\bf r-s}|^2 /r^2 =
  \omega_A \cdot (1 - 2xs_0 /r^2) + O(s^2) \,,
  \end{eqnarray}
i.e., the periodic order parameter is modulated by a
slowly decreasing function. The Abrikosov $\beta$ of this
defect times the volume, $\lim_{V\to \infty} (\beta-\beta_0)V$,
diverges and so the defect energy diverges.

  This unphysical divergence of defect energies of the
vortex lattice is removed when the influence of the
nonlinear GL terms on the solutions $\omega(x,y)$ and
$B(x,y)$ is accounted for. This calculation was performed
in a series of 4 papers \cite{11}: Parts 1 and 2 deal with
the linear elastic energy of the vortex lattice at low
and high inductions $\bar B$. Parts 3 and 4 derive the GL
solutions for the distorted vortex lattice when the vortex
lines are straight and parallel or arbitrarily curved.
The essential result is that the long-ranging modulation
factors like $(1 -2s/r)$ in the linearized solution (14)
become exponentially damped over a new length
$\xi' = 1/k_\psi = \xi/\sqrt{2(1-b)}$ with $b=\bar B /B_{c2}$.
As $b \to 1$, this screening length becomes infinite and the
linearized solution (14) is recovered. At $b < 1$, the
distorted-lattice solution (14) should be replaced by
  \begin{eqnarray}  
  \omega({\bf r}) = \omega_A({\bf r}) \Big[ 1 + \sum_\nu
  {\bf s}_\nu \nabla K_0( |{\bf r -R}_\nu | k_\psi) \Big]^2
                             + O(s^2) \,,
  \end{eqnarray}
where $K_0(x)$ is a modified Bessel function with the limits
$K_0(x) \approx -\ln x$ ($x \ll 1$), $K_0(x) \approx
(\pi /2 x)^{1/2} \, e^{-x}$ ($x \gg 1$). This generalized
expression up to terms linear in the vortex shifts
${\bf s}_\nu$ reproduces the linearized solution (14), (20)
when $k_\psi \to 0$, but it does not possess the correct
zeros at ${\bf r}_\nu = {\bf R}_\nu +{\bf s}_\nu$. This may
be corrected by replacing in (21) the periodic order parameter
$\omega_A({\bf r})$ by the ``phase modulated''
$\omega_A[{\bf r - s(r)}]$ and cutting the infinity of
$K_0$ off. The resulting solution is still
exact up to linear terms in ${\bf s}_\nu$ since
$\nabla \omega_A({\bf r})$ vanishes at the ${\bf R}_\nu$ and
thus the expansion of $\omega_A[{\bf r - s(r)}]$ contains no
linear term.

   The screening length $\xi' = 1/k_\psi$ may be derived by
considering only one Fourier component of the displacement
field,
  \begin{eqnarray}  
  {\bf s}_\nu = {\rm Re} \{ {\bf s}_0 \exp(i{\bf k R}_\nu)\}
  \end{eqnarray}
with ${\bf k} = (k_x, k_y, 0)$ and Re = real part. One may
then write the linearized solution as
  \begin{eqnarray}  
  \omega({\bf r}) = \omega_A({\bf r}) [ 1 +
  {\textstyle{ 1\over 2}} \eta({\bf r}) ]^2  + O(s^2) \,, \\
  \eta({\bf r}) = 2 \sum_\nu {\bf s}_\nu {{\bf r-R}_\nu
  \over ( {\bf r-R}_\nu )^2 }   
  = {2b\over \xi^2} {\rm Re} \Bigg\{ {\bf s}_0 \sum_{\bf K}
  { i({\bf k+K}) \over ({\bf k+K})^2 }
  \exp [ i{\bf (k+K)r} ] \Bigg \} \,.
  \end{eqnarray}
In $\eta({\bf r})$ (24) the terms with reciprocal lattice
vectors ${\bf K} \ne 0$ shift the zeros of $\omega({\bf r})$
(``phase modulation''), while the term ${\bf K}=0$ yields
an ``amplitude modulation'' of $\omega({\bf r})$. This term
diverges as $1/k^2$, i.e., it yields a diverging amplitude
modulation when the wavelength of the displacement field is
large.

  From physical reasons it is clear that this term
$\propto 1/k^2$ has to be cut off, e.g., replaced by
$ 1/(k^2 + k_\psi^2)$. Accounting for all the GL terms
nonlinear in $\omega \propto 1-b$ ($b= \bar B/B_{c2}$,
terms like $\omega^2$, $B^2$, $Q^2$) indeed yields such
a cut off, with $k_\psi^2 = 2(1-b)/\xi^2$. The resulting
solution for periodic ${\bf s(r)}$ may be written as
  \begin{eqnarray}  
  \omega({\bf r}) = \omega_A[ {\bf r -s(r)}] \Bigg[ 1 +
  {2b\over \xi^2} {\nabla {\bf s(r)} \over k^2 +k_\psi^2} \Bigg]
     + O(s^2) \,.
  \end{eqnarray}
In a similar way, the solution for the induction $B(x,y)$ of
the linearized GL theory,
  \begin{eqnarray}  
  B({\bf r}) = \bar B + B_{c2} { \langle \omega \rangle -
  \omega({\bf r})  \over 2\kappa^2 }
  \end{eqnarray}
is modified by the nonlinear terms to give for periodic
${\bf s(r)}$
  \begin{eqnarray}  
  B({\bf r}) = B_0 [ {\bf r -s(r)}] - { \bar B~ \nabla
  {\bf s(r)} \over 1 +k^2 /k_h^2 } + O(s^2)
  \end{eqnarray}
with $k_h^2 = 1/\lambda'^2 = \langle \omega \rangle
 /\lambda^2 \approx (1-b)/\lambda^2$ and $B_0(x,y)$ the ideal
periodic solution for ${\bf s} \equiv 0$. In deriving (27)
all terms containing $k_\psi$ have cancelled. From the
solutions (25) and (27) for periodic ${\bf s(r)}$, the
generalization to arbitrary displacement fields is obtained
by Fourier transform.

\section{Curved vortices}  

  The above method can be extended to 3D displacement fields
${\bf s}_\nu(z) = [s_{\nu x}(z), s_{\nu y}(z), 0]$ describing
distorted lattices of curved vortices,
  \begin{eqnarray}  
  {\bf s}_\nu(z) = \int_{\rm BZ}{d^3 k\over 8\pi^3 n} \,
  {\bf \tilde s(k)} \exp( i {\bf k R}_\nu) \,,  \nonumber \\
  {\bf \tilde s(k)} =\sum_\nu \int\!\! dz\, {\bf s}_\nu(z)
  \exp(-i {\bf k R}_\nu) \,,
  \end{eqnarray}
where now ${\bf r} = (x,y,z)$, ${\bf R}_\nu =(x_\nu, y_\nu, z)$,
$n=B/\Phi_0$, and the ${\bf k}$ integration extends over the
first Brillouin zone of the ideal vortex lattice
[since ${\bf \tilde s (k+K) = \tilde s(k)}$] and over
$-\infty < k_z < \infty$. The coordinate $z$ plays here the
role of a line parameter. The order parameter which solves the
GL equations near $B_{c2}$ and has zeros at the vortex positions
${\bf r}_\nu(z) = {\bf R}_\nu +{\bf s}_\nu(z)$ is
  \begin{eqnarray}  
  \omega({\bf r}) = \omega_A( {\bf r}) \Bigg[ 1 +
  \sum_\nu \int\!\! dz' {\bf s}_\nu(z') \nabla {\exp (
  -| {\bf r-R}'_\nu | k_\psi) \over 2| {\bf r-R}'_\nu |} \Bigg]
      + O(s^2) \,.
  \end{eqnarray}
The 3D solution for the induction ${\bf B(r)} = \nabla\times
{\bf A(r)}$ for periodic ${\bf s(r)}$ after averaging over a
vortex cell may be written as
  \begin{eqnarray}  
  {\bf B(r)} = {\bf \hat z} \bar B + \bar B\, {{\bf\hat z} \nabla
  {\bf s(r)} + \partial {\bf s(r)} /\partial z \over
       1 +k^2/k_h^2 } + O(s^2) \,,  \\
  {\bf A(r)} = {1 \over 2} \bar B\, {\bf\hat z}\times {\bf r}
  + \bar B\, { {\bf s(r)}\times
  {\bf\hat z} \over  1 +k^2/k_h^2 } + O(s^2) \,.
  \end{eqnarray}
These expressions coincide with the first-order expansion
terms (in $s$) of a linear superposition of spherical
``source fields'' centered at each vortex element:
  \begin{eqnarray}  
  {\bf B(r)} = \Phi_0 k_h^2 \sum_\nu \int\! d{\bf r}_\nu
  { \exp (-\rho k_h) \over 4\pi \rho } \,,
  \end{eqnarray}
where $\rho \approx [ ({\bf r-r}_\nu)^2 +a^2 /4 ]^{1/2}$
has an inner cut-off $\approx a/2$, half the vortex spacing
in our derivation from the vortex lattice. The expression
(32) is also the solution of London theory for arbitrarily
arranged curved or straight vortices if one puts
$k_h = 1/\lambda$ (i.e. $b \to 0$) and the vortex core
radius $r_c \approx \xi$ for the inner cutoff. The line
element of the path integral in (32) may be parameterized
with $z$  as line parameter and integration variable,
  \begin{eqnarray}  
   d{\bf r}_\nu = {d{\bf r}_\nu(z) \over dz} dz =
   \Big( {\bf\hat z} + {d {\bf s}_\nu(z) \over dz} \Big) dz \,.
  \end{eqnarray}

\section{Nonlocal elasticity of the vortex lattice}  

  The distorted-lattice solution up to terms linear in the
displacements ${\bf s}$ can be used to calculate the linear
elastic energy of the vortex lattice,
$F_{\rm elast} =F\{ {\bf s}_\nu \} -F\{ {\bf s}_\nu \equiv 0 \}$,
referred to the perfect lattice (the equilibrium state).
The most general expression quadratic in the 2D displacements
${\bf s}_\nu(z)$, or in their Fourier transforms
${\bf \tilde s(k)} = (\tilde s_x, \tilde s_y, 0)$, (28), is
  \begin{eqnarray}  
   F_{\rm elast} ={1\over 2} \int_{\rm BZ} {d^3 k\over 8\pi^3 n}
   \, \tilde s_\alpha({\bf k}) \Phi_{\alpha\beta}({\bf k})
      \tilde s_\beta({\bf -k}) \,,
  \end{eqnarray}
where the sum over the indices $\alpha$, $\beta$ = ($x,y$) is
taken. The $2 \times 2$ matrix $\Phi_{\alpha \beta}$ is the
elastic matrix. This expression applies for both an elastic
continuum and for a lattice. For a lattice $\Phi_{\alpha \beta}$
is periodic, $\Phi_{\alpha \beta} ({\bf k+K}) =
\Phi_{\alpha \beta}({\bf k})$, and thus the integral should be
restricted to the first Brillouin zone (BZ). The BZ for the
triangular lattice is a hexagon, and for the square lattice a
square. Where required, the BZ may be approximated by a circle
with radius $k_B = (2b)^{1/2} / \xi$, $b=\bar B/B_{c2}$, and
area $\pi k_B ^2 = 4\pi^2 n$, $n = \bar B /\Phi_0 $.

  For a uniaxial elastic continuum the elastic matrix
$\Phi_{\alpha \beta}(k_x, k_y, k_z)$ has the form
  \begin{eqnarray}  
  n \Phi_{\alpha \beta}({\bf k}) = (c_{11} -c_{66})
  k_\alpha k_\beta +\delta_{\alpha\beta} [( k_x^2 +k_y^2 )
  c_{66} + k_z^2 c_{44} ]\,.
  \end{eqnarray}
In it the coefficients are the elastic moduli: $c_{11}-c_{66}$
the isotropic compression modulus, $c_{11}$ the uniaxial
compression modulus, $c_{66}$ the shear modulus, and $c_{44}$
the tilt modulus. The elastic moduli of the vortex lattice
are obtained by deriving the elastic energy, e.g., from GL
theory and comparing it at $k_x^2+k_y^2 \ll k_B^2$ with the
continuum limit (35). This yields
  \begin{eqnarray}  
  c_{11}(k) &=& {\bar B^2 \over \mu_0} {\partial B_a \over\partial
    \bar B }\, {1 \over (1 +k^2/k_h^2) (1 +k^2/k_\psi^2) }
    + c_{66} \\
  c_{66} &=& {\bar B B_{c2} \over 8 \kappa^2 \mu_0}\,
      (1-b)^2 {(2\kappa^2 -1) 2\kappa^2 \over
    [2\kappa^2 -1 +1/\beta_A]^2 } (1-0.3b) \\
  c_{44}(k) &=& {\bar B^2 \over \mu_0}\, {1 \over 1 +k^2/k_h^2}
    + { \bar B (B_a -\bar B) \over \mu_0} \,.
   \end{eqnarray}
These expressions are exact at large reduced induction
$b = \bar B /B_{c2} \to 1$ and for all $\kappa$, but
they are written such that they reduce to the correct values
also in the limit of small induction $\bar B \ll B_{c2}$.
In $c_{66}$, $\beta_A = 1.160$ is the Abrikosov parameter
of the triangular lattice (the square lattice is unstable
and thus has negative $c_{66}$); the third
factor reduces to
 1 for $2\kappa^2 \gg 1$ and to
 $(2\kappa^2-1)\beta_A^2 \to 0$ for $\kappa \to 1/\sqrt 2$,
which means the shear stiffness of the vortex lattice
is zero in superconductors with $\kappa = 0.71$;
the factor $1 -0.3 b$ interpolates between the
the correct limits at $b \to 1$ and $b \to 0$. In
particular, for $b\ll 1$ and $2\kappa^2 \gg 1$, (37)
reproduces the London result
$c_{66} = \bar B B_{c2} /(8 \kappa^2 \mu_0)$.

  An interesting result is the dependence
of $c_{11}$ (36) and $c_{44}$ (38) on $k = |{\bf k}|$,
which means the elasticity of the vortex lattice is
{\bf non-local}. In the limit of
uniform stress, $k \to 0$, these expressions
reproduce the known values of the compression and tilt
moduli obtained by thermodynamics, $c_{11} -c_{66}
 = (\bar B^2 / \mu_0) \partial Ba / \partial \bar B$,
$c_{44} =\bar B B_a / \mu_0$. However, when the wavelength
of the periodic compression or tilt decreases, i.e., the
wave vector $k$ increases, these moduli decrease. This
means, the vortex lattice is softer for short-wavelengths
compression and tilt than it is for long wavelengths.
The two characteristic lengths or wave vectors were already
introduced above,
$k_h = 1/\lambda' \approx \sqrt{1-b}/\lambda$ and
$k_\psi = 1/\xi' = \sqrt{2(1-b)}/\xi$.

  This dispersion or elastic non-locality means, e.g.,
that a point force exerted by a small pinning center on
the vortex lattice, deforms the vortex on which it acts
not like plugging a string but more, causing a sharp cusp
since a local deformation costs little energy. If the
interaction of the vortices with the pinning center is
via the order parameter $|\psi|^2$ or via the gradient
term in the GL functional, then this interaction itself
is nonlocal, smeared over the length $\xi' =1/k_\psi$.
In the expressions for the elastic force and the
elastic energy there is thus a factor $1 +k^2/k_\psi^2$
in the numerator that compensates the same factor in
the denominator originating from $c_{11}(k)$, (36).
Therefore, the factor $1/(1 +k^2/k_\psi^2)$ in $c_{11}$
has no physical meaning in pinning problems since
near $B_{c2}$ where $\xi'$ can be larger than the
vortex spacing $a$, it is not possible to exert a
pinning force on one single zero of the order parameter
but only on an area with radius $\xi'$ containing several
such zeros. The nonlocality factor $1/(1 +k^2/k_h^2)$
in $c_{44}$, however, is important in pinning theories
since it strongly enhances the elastic deformations
caused by small pins acting on the vortex cores.
In (not very realistic) models where the pinning force
acts only the magnetic field of the vortex but not on
the vortex cores, this enhancement of the elastic
displacement may vanish, cancelled by the non-locality
of this model force.

The correct, non-local elasticity thus effectively
softens the vortex lattice and leads to large,
pinning-caused distortions and disorder of the vortex
lattice. Furthermore, the thermal fluctuations of the
vortex lattice are strongly enhanced by this non-local
elasticity. In both cases the lattice softening is
caused mainly by the dispersion of $c_{44}(k)$, while
the dispersion and reduction of $c_{11}(k)$ is
not so important since the shear modulus $c_{66}$ is
typically much smaller than $c_{11}(k)$ and the shear
modes of the elastic deformation thus dominate over
the compressional modes.

\begin{figure}[tbh]  
\epsfxsize= .7\hsize  \vskip 1.0\baselineskip \centerline{
\epsffile{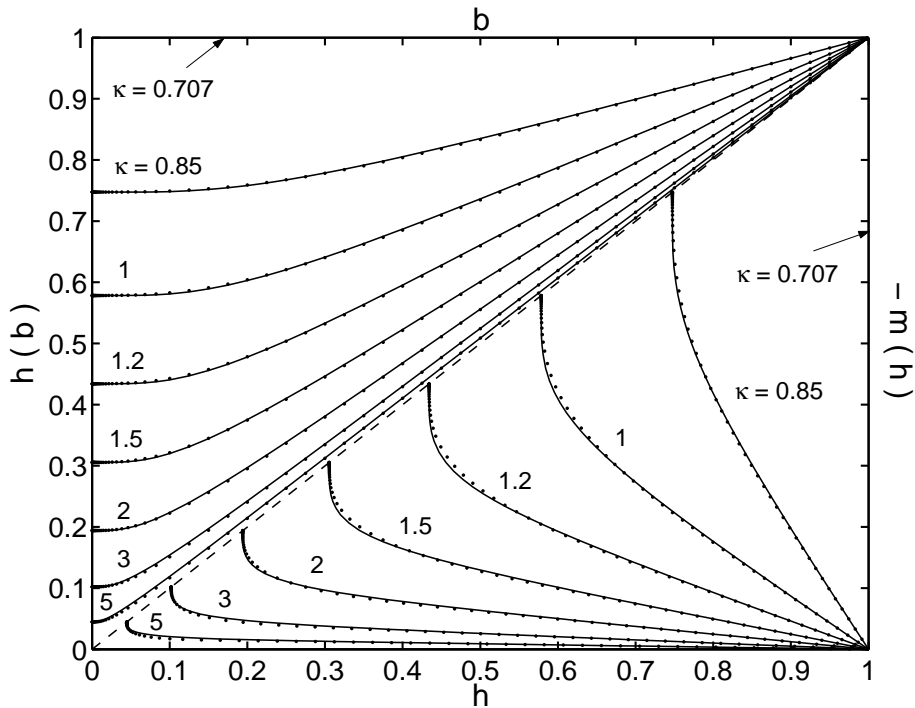}}  \vspace{ 0 mm}  
\caption{\label{fig3}
The magnetization curves of the triangular vortex lattice
(solid lines, numerical result), coinciding within line thickness
with those of the square lattice. Shown are $h = B_a/B_{c2}$ versus
$b=\bar B/B_{c2}$ (upper left triangle) and $-m = h-b$ versus
$h$ (lower right triangle). The dots show the fit, Eq.~(59),
good for $\kappa \le 20$.
 } \end{figure}   

\begin{figure}[tbh]  
\epsfxsize= .7\hsize  \vskip 1.0\baselineskip \centerline{
\epsffile{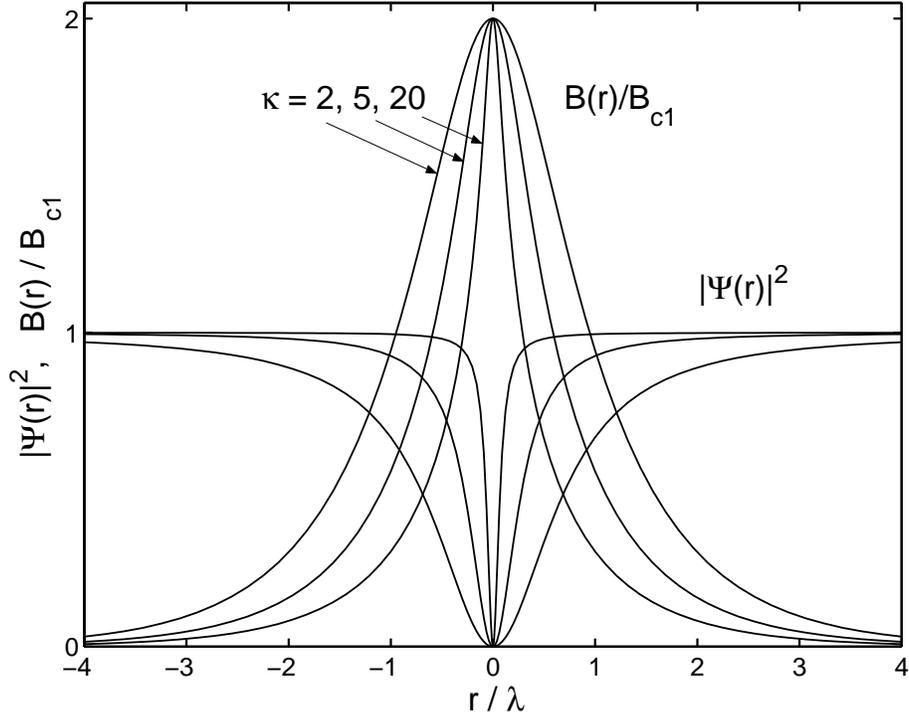}}  \vspace{ 0 mm}  
\caption{\label{fig4}
The magnetic field $B(r)$ and order parameter $|\psi(r)|^2$ of
an isolated vortex line calculated from Ginzburg-Landau theory
for GL parameters $\kappa$ = 2, 5, and 20. For such large
$\kappa$ the field in the vortex center is twice the applied
equilibrium field, $B(0) \approx 2 B_{c1} = 2 B_a$.
 } \end{figure}   

\begin{figure}[tbh]  
\epsfxsize= .7\hsize  \vskip 1.0\baselineskip \centerline{
\epsffile{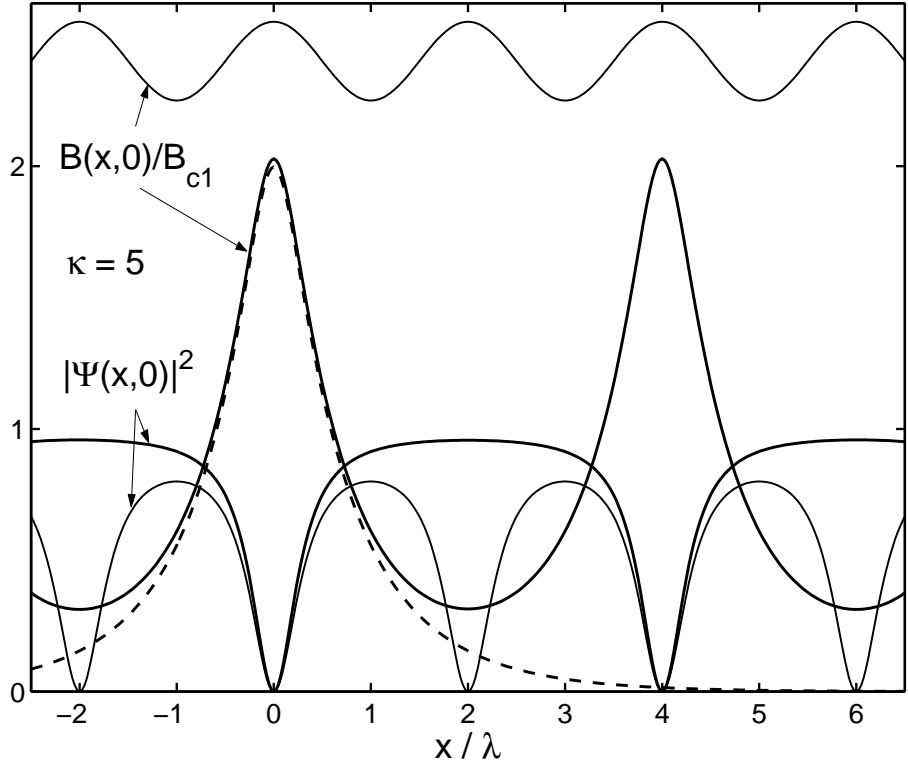}}  \vspace{ 0 mm}  
\caption{\label{fig5}
 Two profiles of the magnetic field $B(x,y)$ and
order parameter $|\psi(x,y)|^2$ along the $x$ axis
(nearest neighbor direction) for triangular vortex lattices
with lattice spacings $a=4\lambda$ ($b=0.073$, bold lines) and
$a=2\lambda$ ($b=0.018$, thin lines). The dashed line shows the
magnetic field of the isolated flux line from Fig.~4.
From Ginzburg-Landau theory for $\kappa=5$.
 } \end{figure}   

\begin{figure}[tbh]  
\epsfxsize= .7\hsize  \vskip 1.0\baselineskip \centerline{
\epsffile{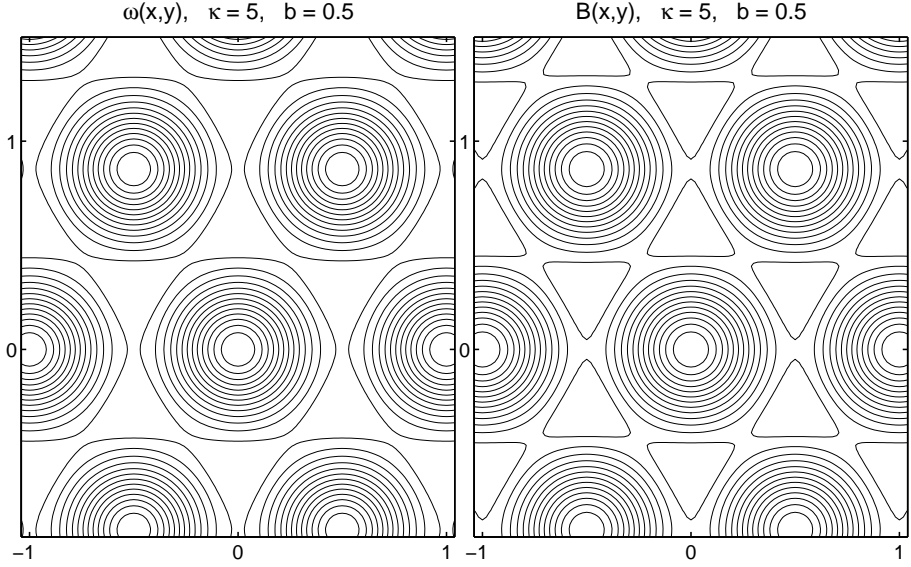}}  \vspace{ 0 mm}  
\caption{\label{fig6}
Countour lines of $\omega(x,y) = |\psi|^2$ and $B(x,y)$ for
$\kappa=5$, $b=0.5$.
 } \end{figure}   

\begin{figure}[tbh]  
\epsfxsize= .7\hsize  \vskip 1.0\baselineskip \centerline{
\epsffile{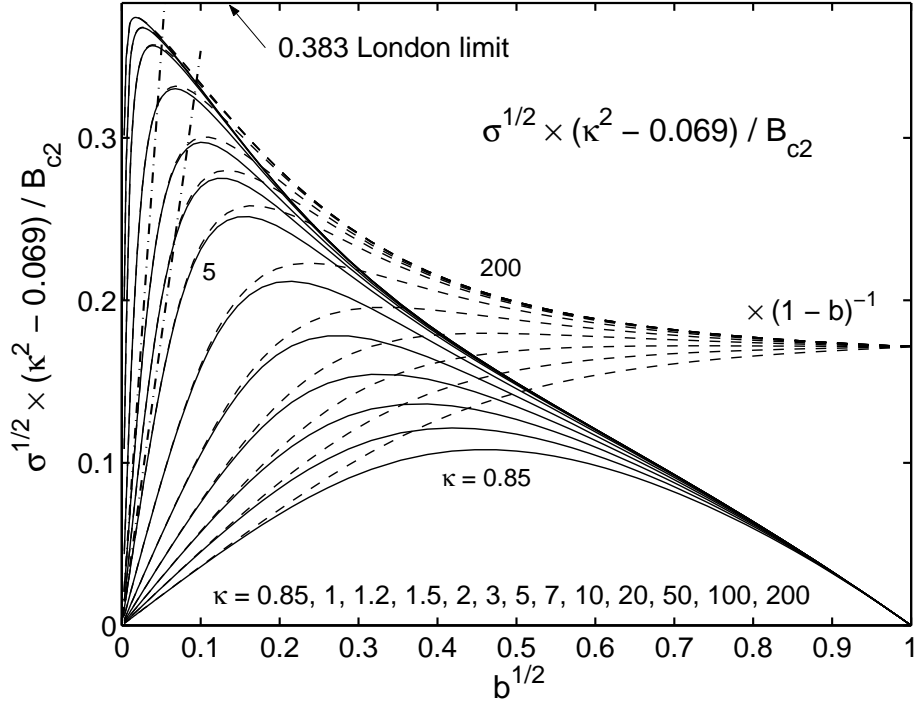}}  \vspace{ 0 mm}  
\caption{\label{fig7}
The magnetic field variance
$\sigma = \langle [B(x,y) - B]^2 \rangle$ of the triangular FLL
for $\kappa=0.85$, 1, $\dots$, 200
plotted in units of $B_{c2}$ as
$\sqrt{\sigma}\cdot(\kappa^2-0.069)/B_{c2}$ (solid lines) such
that the curves for all $\kappa$ collapse near $b=1$.
The dashed lines show the same functions divided by $(1-b)$
such that they tend to a finite constant 0.172 at $b=1$. All
curves are plotted versus $\sqrt b = \sqrt{\bar B/B_{c2}}$ to
stretch them at small $b$ values and show that they go to
zero linearly. The upper frame 0.383 is the usual London
approximation. The limit for very small $b$ is shown as two
dash-dotted straight lines for $\kappa=5$ and $\kappa =10$.
The upper frame 0.383 shows the usual London approximation.
 } \end{figure}   

\begin{figure}[tbh]  
\epsfxsize= .7\hsize  \vskip 1.0\baselineskip \centerline{
\epsffile{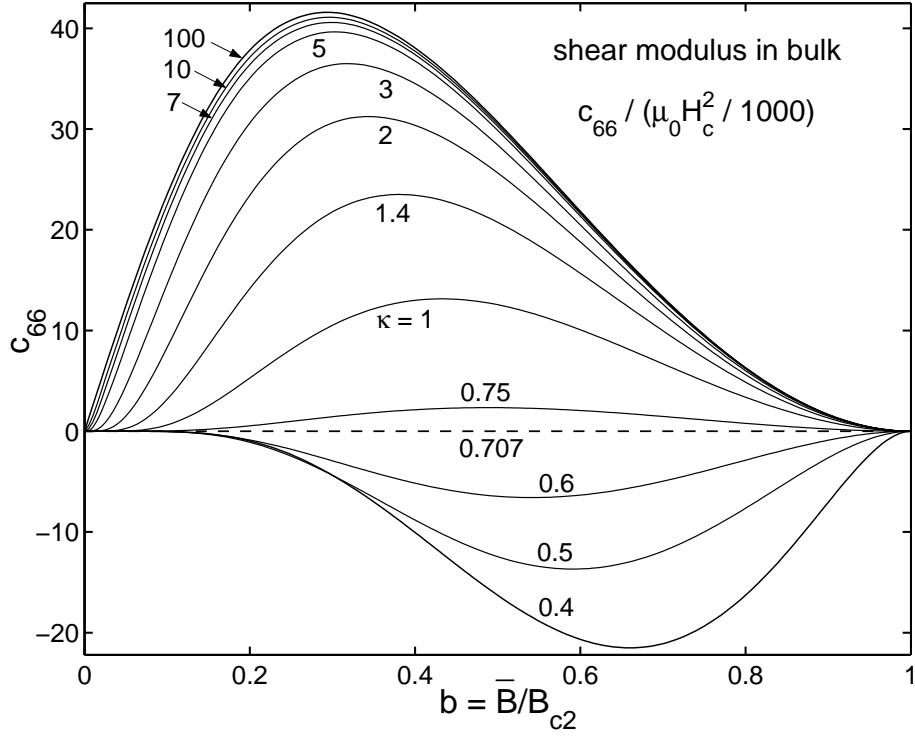}}  \vspace{ 0 mm}  
\caption{\label{fig8}
The shear modulus $c_{66}$ of the triangular vortex
lattice in bulk superconductors as function of the reduced
induction $b=\bar B / B_{c2}$
for GL parameters $\kappa = 0.4$, 0.5, 0.6, .707, 0.75, 1,
1.4, 2, 3, 5, 7, 10, 100, in units $B_c^2 / (1000 \mu_0)$.
For $\kappa < 2^{-1/2} = 0.707$ one formally has negative shear
modulus $c_{66} < 0$, though vortices and a vortex lattice are
energetically not favorable in bulk type-I superconductors.
 } \end{figure}   

\begin{figure}[tbh]  
\epsfxsize= .7\hsize  \vskip 1.0\baselineskip \centerline{
\epsffile{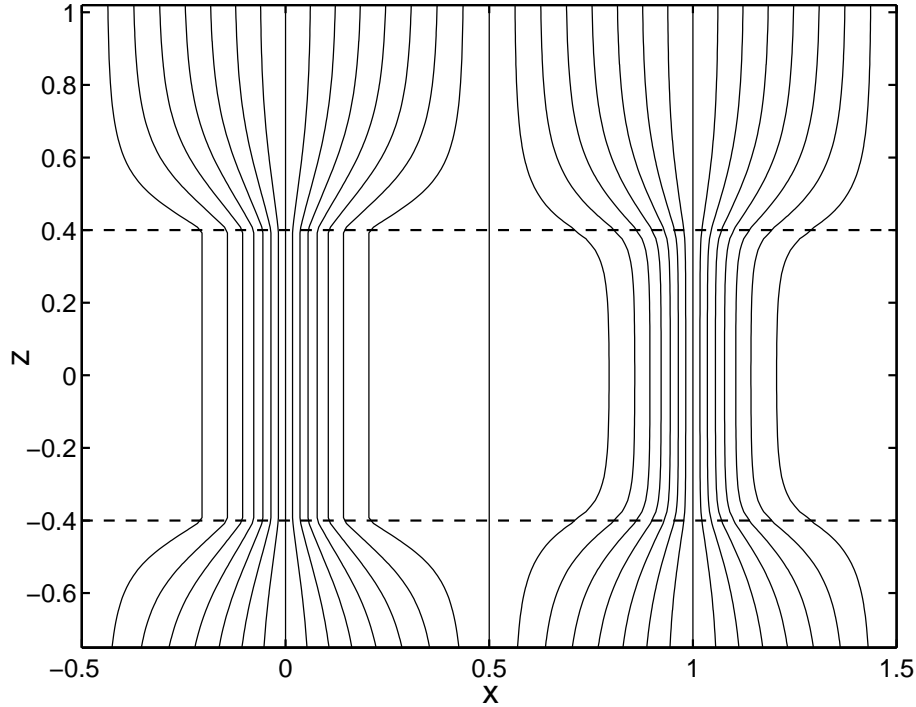}}  \vspace{ 0 mm}  
\caption{\label{fig9}
Magnetic field lines  for a superconductor film calculated from
Ginzburg-Landau theory for the triangular vortex lattice.
Shown is the example
$b=\bar B/B_{c2} = 0.04$, $\kappa = 1.4$, triangular lattice
with vortex spacing (unit length) $x_1 =3^{-1/4}(2\Phi_0
  / \bar B)^{1/2} =5x_1(B_{c2}) \approx 10 \lambda$,
film thickness $d = 0.8 x_1 \approx 8 \lambda$.
The left half shows the field lines that would apply if
the field inside the film would not change near the surfaces
$z=\pm d/2$ marked by dashed lines. The right half shows the
correct solution. The density of the depicted field lines is
proportional to $|{\bf B(r)}|$.
 } \end{figure}   

\begin{figure}[tbh]  
\epsfxsize= .7\hsize  \vskip 1.0\baselineskip \centerline{
\epsffile{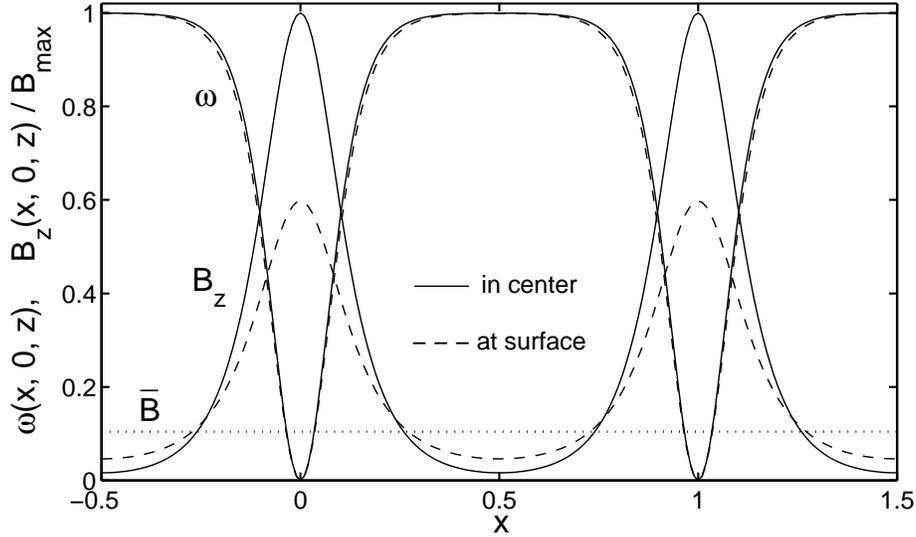}}  \vspace{ 0 mm}  
\caption{\label{fig10}
Profiles of order parameter $\omega(x,0,z)$ and magnetic field
$B_z(x,0,z)$ for the case of Fig.~9, film thickness
$d = 0.8 x_1 \approx 8 \lambda$.
The solid lines show $\omega$ and $B$ in the center
of the film ($z=0$) and the dashed lines at the film surfaces.
The dotted line indicates the average induction $\bar B$
equal to the applied field $B_a$.
 } \end{figure}   

\begin{figure}[tbh]  
\epsfxsize= .7\hsize  \vskip 1.0\baselineskip \centerline{
\epsffile{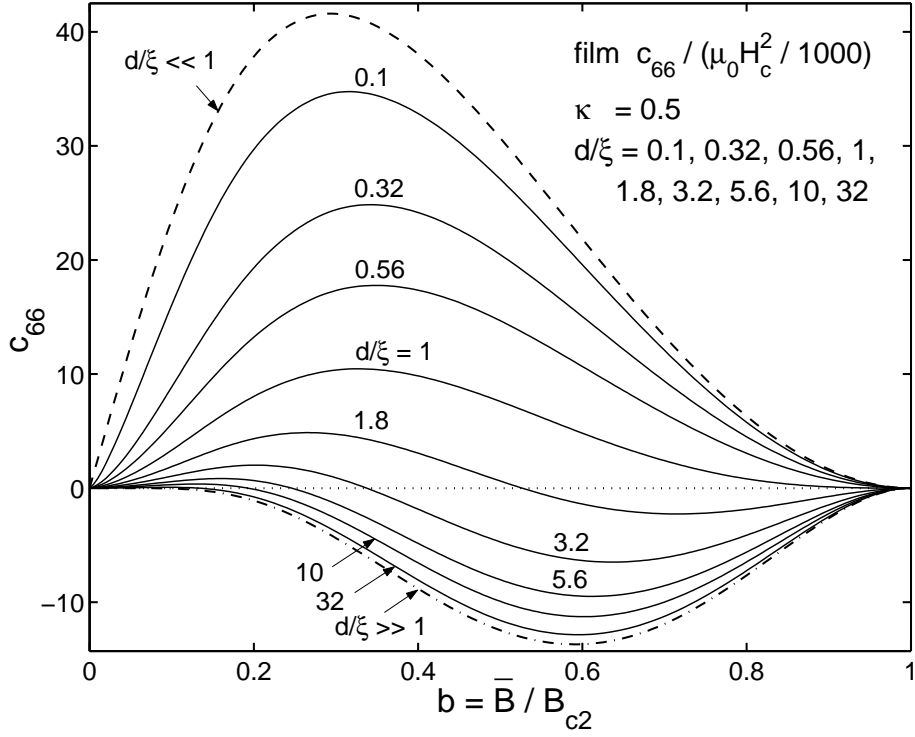}}  \vspace{ 0 mm}  
\caption{\label{fig11}
The shear modulus $c_{66}$ of the triangular vortex
lattice in films with thicknesses $d/\xi = 0.1$, 0.32, 0.56,
1, 1.8, 3.2, 5.6, 10, and 32, plotted versus $b$ for
$\kappa = 0.5$. This $c_{66}$ is positive, i.e., the triangular
vortex lattice is stable, for sufficiently thin films or for
small inductions. For $d \gg \xi$ the bulk $c_{66}$ at the
same $\kappa =0.5$ is reached (dash-dotted line), and for
$d \ll \xi$ the bulk $c_{66}$ in the limit $\kappa \gg 1$
is reached (dashed line).
 } \end{figure}   

\begin{figure}  
\epsfxsize= .7\hsize  \vskip 1.0\baselineskip \centerline{
\epsffile{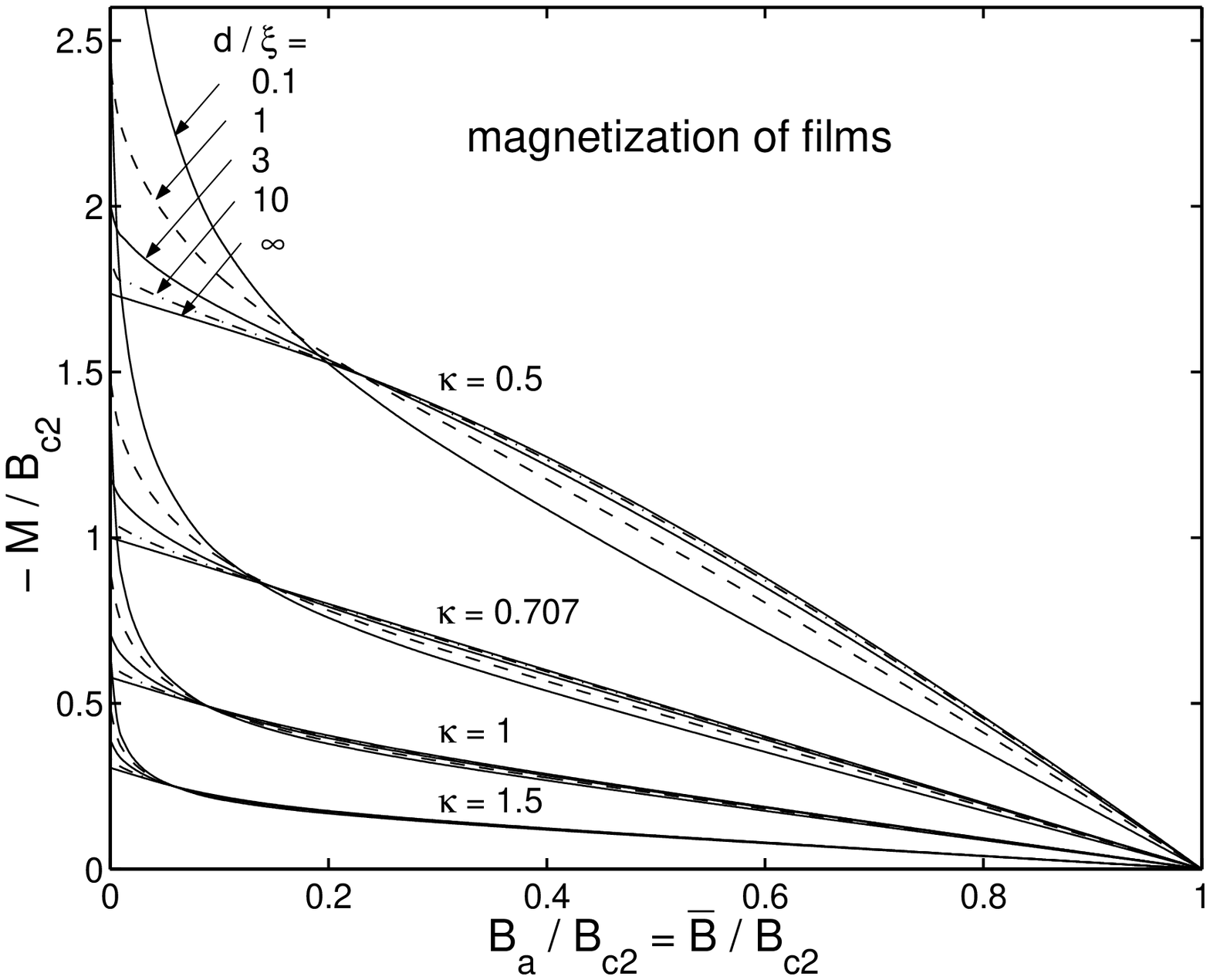}}  \vspace{.1cm}
\caption{\label{fig12}
The magnetization $M$ of infinite films of thickness
$d/\xi = 0.1$, 1 ,3, 10, $\infty$ with a triangular
vortex lattice generated by a perpendicular magnetic field
$B_a$. Plotted is $-M / B_{c2}$ versus $b =\bar B/B_{c2}
= h= B_a/B_{c2}$ for $\kappa = 0.5$, 0.707, 1, 1.5.
 } \end{figure}   

\section{Vortex arrangements at low inductions}  

  At low inductions $b < 0.2$ and not too small $\kappa >2$,
the GL theory for arbitrary 3D arrangements of vortices reduces
to the London theory, which may be expressed by the energy
functional
  \begin{eqnarray}  
   F\{ {\bf B} \} ={\mu_0 \over 2} \int\! d^3 r [ B^2
   + \lambda^2 (\nabla \times {\bf B})^2 ] \,.
  \end{eqnarray}
Here $\lambda$ is the London depth equal to the GL magnetic
penetration depth. Minimizing $F\{ {\bf B} \}$ with respect
to the induction ${\bf B(r)}$ using $\nabla {\bf B} =0$,
and adding appropriate singularities along the positions
${\bf r}_\nu(z)$ of the vortex cores, one obtains the
modified London equation \cite{12},
  \begin{eqnarray}  
   ( -\lambda^2 \nabla^2 +1) {\bf B(r)} = \Phi_0 \sum_\nu
   \int \! d{\bf r}_\nu \, \delta_3({\bf r-r}_\nu)
  \end{eqnarray}
with $\delta_3$ the 3D delta function. From this one obtains
the energy of an arbitrary arrangement of straight or
curved vortices,
  \begin{eqnarray}  
   F\{ {\bf r}_\nu(z) \} = {\Phi_0^2 \over 8\pi\lambda^2
   \mu_0 } \sum_\mu \sum_\nu \int\!\! d{\bf r}_\mu \!
   \int\!\! d{\bf r}_\nu { \exp(-r_{\mu\nu} /\lambda )
   \over r_{\mu\nu} } \,.
  \end{eqnarray}
In this double sum the terms $\mu \ne \nu$ describe the
pairwise interaction of the vortex line elements
$d{\bf r}_\mu$, $d{\bf r}_\nu$ over the distance
$r_{\mu\nu} = |{\bf r}_\mu - {\bf r}_\nu|$. The term
$\mu = \nu$  is the self-energy of the $\mu$th vortex
line, which depends on the shape of this vortex. In it
an inner cut off is needed, obtained, e.g., by putting
$r_{\mu\mu}^2(z,z') = |{\bf r}_\mu(z) - {\bf r}_\mu(z')|^2
 + r_c^2$ with $r_c \approx \xi$ the vortex core radius,
to avoid divergence when in the integral the parameters
equal, $z = z'$.

  From the GL nonlocal elastic energy  (34)-(38) one
may construct an effective interaction potential between
vortex line elements such that the full nonlocal linear
elastic energy is reproduced at small displacements \cite{13}.
At the same time, this interaction at low $b \ll 1$
reproduces the London interaction for arbitrary vortex
arrangements, and an approximate GL interaction valid at
all $b$ and $\kappa$,
  \begin{eqnarray}  
   F\{ {\bf r}_\nu(z) \} = {\Phi_0^2 \over
   8\pi \lambda'^2 \mu_0 } \sum_\mu \sum_\nu \!\Bigg[ \!
   \int\!\!  d{\bf r}_\mu \! \int\!\! d{\bf r}_\nu
   { \exp(-r_{\mu\nu} /\lambda')
    \over  r_{\mu\nu} } -
   \int\!\!  |d{\bf r}_\mu| \! \int\!\! |d{\bf r}_\nu|
   { \exp(-r_{\mu\nu} /\xi' )
     \over  r_{\mu\nu} } \, \Bigg]
  \end{eqnarray}
with $r_{\mu\nu} = |{\bf r}_\mu - {\bf r}_\nu|$. For
$b \ll 1$ the first term in (42) reproduces the magnetic
repulsion of London vortices with
$\lambda' = \lambda /\sqrt{1-b} \approx \lambda$;
this magnetic interaction is vectorial due to the
product $d{\bf r}_\mu \cdot d{\bf r}_\nu$ containing
the cosine of the angle between two line elements. The
second term of shorter range $\xi' = \xi/\sqrt{2(1-b)}$
may be interpreted as an attraction caused by the overlap
of the vortex cores, the regions where the order parameter
is reduced: two overlapping cores require less (positive)
condensation energy than two separated cores, thus the
cores attract. This attraction has scalar character,
hence the product $|d{\bf r}_\mu|\, |d{\bf r}_\nu|$.
The attractive second term in (42) removes the logarithmic
divergence of the magnetic repulsion at zero distance,
since both terms have the same singularity but of
opposite sign, such that they cancel. The total
potential (42) is thus a smooth function that at
$r_{\mu\nu} = 0$ starts with a finite value and then
decreases monotonically to zero with increasing distance
$r_{\mu\nu} \to \infty$.

  For straight parallel vortex lines the general 3D
energy expression (42) simplifies to the sum of the
vortex self energies,
$F_{\rm self} = \Phi_0 B_{c1} /\mu_0$ per unit length
and per vortex, and the interaction energy $F_{\rm int}$
of all vortices per unit length,
  \begin{eqnarray}  
   F_{\rm int}\{ {\bf r}_\nu \} = {\Phi_0^2 \over
   2\pi \lambda'^2 \mu_0 } \sum_\mu \sum_{\nu >\mu}
   \!\Bigg[ K_0 \Bigg( { |{\bf r}_\mu -{\bf r}_\nu|
     \over \lambda'} \Bigg) - K_0 \Bigg({ |{\bf r}_\mu
     -{\bf r}_\nu| \over \xi'} \Bigg) \Bigg] \,.
  \end{eqnarray}
Here $K_0(x)$ is a modified Bessel function, see Eq.~(21).
The effective 2D interaction potential in (43) is a smooth,
monotonically decreasing function with a finite value at
$r_{\mu\nu} = 0$ since the two logarithmic singularities
of the $K_0$ functions cancel each other. As in the 3D
expression (42), the first term in (43) is the magnetic
repulsion of the straight vortices, and the second term is
an attraction due to gain in condensation energy during
the overlap of vortex cores.

\section{Vortex lattice solution for all $\kappa$ and $\bar B$} 

   Abrikosov's solution method for the periodic vortex lattice
starts from the linearized GL theory and is thus valid only
at large inductions $\bar B$ near the upper critical field
$B_{c2}$. First numerical solutions for all $\bar B$ and $\kappa$
were obtained by the ``circular cell method'' \cite{14} that
approximates the hexagonal Wigner-Seitz cell of the triangular
vortex lattice by a circle and solves a cylindrically symmetric
problem, see also Ref.~\cite{15}.
The periodic solution in the entire ranges of reduced induction
$0 < b=\bar B/B_{c2} < 1$ and GL parameter
$1/\sqrt2 \le \kappa < \infty$ may be obtained for bulk
superconductors by the following numerical method \cite{15,16,17}.
We start from the free
energy functional $F$, Eq.~(2), and minimize it with respect
to the real and periodic functions $\omega(x,y) =f^2 =|\psi|^2$
(order parameter) and ${\bf Q}(x,y)={\bf A}-\nabla\varphi/\kappa$
(negative super velocity) or
${\bf \hat z} B(x,y) =\nabla\times {\bf Q}$
(induction). We consider periodic lattices with one flux quantum
per vortex. In the sense of a Ritz variational method we use
Fourier series for the periodic trial functions with a finite
number of Fourier coefficients $a_{\bf K}$ and $b_{\bf K}$,
  \begin{eqnarray}  
  \omega({\bf r})&=&\sum_{\bf K} a_{\bf K} (1-\cos{\bf K r})\,,\\
  B({\bf r}) &=& \bar B +\sum_{\bf K} b_{\bf K}\,\cos{\bf K r}\,,\\
  {\bf Q(r)} &=& {\bf Q}_A({\bf r}) + \sum_{\bf K} b_{\bf K}
  {{\bf\hat z \times K} \over K^2} \sin{\bf K r} \,,
  \end{eqnarray}
where ${\bf r} = (x,y)$ and ${\bf K} = (K_x,K_y)$
are the reciprocal lattice vectors (8) of the vortex
lattice with positions (7). In all sums here and below the
term ${\bf K} =0$ is excluded. In (46) ${\bf Q}_A(x,y)$ is
the super velocity of the Abrikosov $B_{c2}$ solution, which
satisfies
  \begin{eqnarray}    
  \nabla \times {\bf Q}_A = \Big[ \bar B -\Phi_0 \sum_{\bf R}
  \delta_2({\bf r-R}) \Big] {\bf\hat z}\,,
  \end{eqnarray}
where $\delta_2({\bf r}) = \delta(x)\delta(y)$ is the 2D delta
function. This relation shows that ${\bf Q}_A$ is the velocity
field of a lattice of ideal vortex lines but with zero average
rotation. Close to each vortex center one has
${\bf Q}_A({\bf r}) \approx {\bf r' \times\hat z}/(2\kappa r'^2)$
and $\omega({\bf r}) \propto r'^2$ with ${\bf r'= r-R}$.
In principle ${\bf Q}_A({\bf r})$ may be expressed as a slowly
converging Fourier series by integrating (47) using
${\rm div} {\bf Q} = {\rm div} {\bf Q}_A = 0$ as in
Ref.~\cite{16}. But it is more convenient to take
${\bf Q}_A$ from the exact relation
  \begin{eqnarray}    
  {\bf Q}_A({\bf r}) = {\nabla \omega_A \times {\bf\hat z}
  \over 2\, \kappa\, \omega_A } \,,
  \end{eqnarray}
where $\omega_A(x,y)$ is the Abrikosov $B_{c2}$ solution given
by the rapidly converging series (11). With (48) the numerical
method becomes highly accurate. Note that the ansatz~(46)
assumes that ${\rm div} {\bf Q} =0$. This assumption can be
shown to be exact at high and low inductions, but I did not
find a proof that it is true in the general case, though it is
satisfied numerically with high precision for the periodic
vortex lattice at all $\bar B$ and $\kappa$.

  The solutions $\omega({\bf r})$ and $B({\bf r})$ may be
computed by using a finite number of Fourier coefficients
$a_{\bf K }$ and $b_{\bf K}$ and minimizing the free energy
$F(B, \kappa, a_{\bf K }, b_{\bf K})$ with respect to these
coefficients as done in \cite{16}. However, a much faster and
more accurate solution method \cite{15,17} is to iterate the two
GL equations $\delta F /\delta \omega =0$ and
$\delta F /\delta {\bf Q} =0$ written in appropriate form.
The iteration is stable and converges rapidly if one
isolates a term
$(-\nabla^2 + {\rm const}) (\omega, \, {\bf Q})$ on the
l.h.s.\ and puts the remaining terms to the r.h.s.\ as an
``inhomogeneity'' of such London-like equations, e.g.,
  \begin{eqnarray}    
  (-\nabla^2 +2\kappa^2)\, \omega ~&=&~ 2\kappa^2 (2 \omega
  -\omega^2 -\omega Q^2 -g) \,, ~\\
  (-\nabla^2 +\bar \omega)\, {\bf Q}_b ~&=&~ -\omega{\bf Q}_A
   -(\omega -\bar \omega){\bf Q}_b \,,
  \end{eqnarray}
with the abbreviations
$g({\bf r})=(\nabla\omega)^2 /(4\kappa^2 \omega)$,
${\bf Q}_b = {\bf Q - Q}_A$,
$\nabla\times {\bf Q}_b =B({\bf r}) - \bar B$,
and $\bar\omega =\langle \omega \rangle =\sum'_{\bf K} a_{\bf K}$.
Equations (49),  (50) introduce some ``penetration depths''
$(2\kappa^2)^{-1/2}  = \xi /\sqrt2$  and
$\bar \omega^{-1/2} = \lambda/ \bar\omega^{1/2}$ (in real units),
which stabilize the convergence of the iteration. Acting on the
Fourier series $\omega$ (44) and ${\bf Q}_b$ (46) the Laplacian
operator $\nabla^2$ yields a factor $-K^2$, which facilitates
the inversion of (49) and (50). Using the orthonormality
  \begin{eqnarray}   
  2\, \langle \cos {\bf K r} \cos {\bf K' r} \rangle =
                                  \delta_{\bf K K'}
  \end{eqnarray}
valid for ${\bf K} \ne 0$, one obtains from (44), (45)
$a_{\bf K} = -2\langle \omega({\bf r}) \cos{\bf Kr} \rangle$ and
$b_{\bf K} =  2\langle      B({\bf r}) \cos{\bf Kr} \rangle$.
The convergence of the iteration is considerably improved by
adding a third equation which minimizes $F$~(2) with
respect to the amplitude of $\omega$, i.e.,
$\partial F / \partial \bar \omega = 0$. This step gives the
largest decrease of $F$. The resulting three iteration equations
for the parameters $a_{\bf K}$ and $b_{\bf K}$ then read
  \begin{eqnarray}    
  a_{\bf K} &:=& {4\kappa^2 \langle (\omega^2 +\omega Q^2 -2\omega
    + g ) \cos{\bf K r} \rangle \over K^2 + 2\kappa^2 } \,,
                                                      \\
  a_{\bf K} &:=& a_{\bf K} \cdot \langle \omega -\omega Q^2 -g
    \rangle ~/~ \langle \omega^2 \rangle\,,
                                                       \\
  b_{\bf K} &:=& {-2 \langle [(\omega -\bar\omega)B({\bf r})
    + p\, ] \cos{\bf K r} \rangle \over K^2 + \bar \omega } \,,
  \end{eqnarray}
with $p = (\nabla \omega \times {\bf Q}) {\bf\hat z} =
 Q_y \partial\omega /\partial x - Q_x \partial\omega /\partial y$
and $g=(\nabla\omega)^2 /(4\kappa^2 \omega)=(\nabla f)^2/\kappa^2$
as above.  The solutions $\omega({\bf r})$, ${\bf B(r)}$, and
${\bf Q(r)}$ are then obtained by starting, e.g., with
$a_{\bf K} = (1-b)\, a_{\bf K}^A $ [the Abrikosov solution (11)]
and $b_{\bf K} = 0$ and then iterating the three equations (52)-(54)
by turns until the coefficients do not change any more. After
typically 25 such triple steps, the solution stays constant
to all 15 digits and the GL equations are exactly satisfied.

  Since all terms in (52) - (54) are smooth periodic
functions of ${\bf r}$, high accuracy is achieved by using
a regular spatial 2D grid, e.g.,
$x_i = (i- 1/2)x_1 /N_x$    ($i=1 \dots N_x$) and
$y_j = (j- 1/2)y_2 /(2N_y)$ ($j=1 \dots N_y$,
$2N_y \approx N_x y_2 /x_1$) with constant weights
$x_1/N_x$ and $y_2/(2N_y)$. These $N=N_x N_y$ = 100 to 5000
grid points fill the rectangular basic area $0\le x \le x_1$,
$0\le y \le y_2/2$, which is valid for any unit cell with the
shape of a parallelogram. Spatial averaging $\langle ...\rangle$
then just means summing $N$ terms and dividing by $N$.
  Best accuracy is achieved by considering all ${\bf K}_{mn}$
vectors within a half circle $|{\bf K}_{mn}| \le K_{\rm max}$, with
$K_{\rm max}^2 \approx 20 N/(x_1 y_2)$ chosen such that the number
of the ${\bf K}_{mn}$ is slightly less than the number $N$
of grid points.
 The high precision of this method may be checked with the
identity $B(x,y)/B_{c2} = 1 -\omega(x,y)$, which is valid at
$\kappa = 1/\sqrt 2$ for all $b$. This relation is confirmed
with an error $<10^{-9}$. The reversible magnetization
$M = B -B_a$ and the equilibrium field
$B_a =\mu_0 \partial F/\partial \bar B$ (the applied field)
are easily computed from Doria's virial theorem \cite{18},
which in our reduced units reads
   \begin{equation}   
   B_a ={ \langle f^2 - f^4 + 2\,B(x,y)^2 \rangle \over
                  2\,\langle B \rangle }\,.
   \end{equation}
In this way we find the lower critical field \cite{15},
$B_{c1}(\kappa) = \lim_{\bar B \to 0} B_a(\bar B, \kappa)$,
   \begin{eqnarray}  
   B_{c1}(\kappa) &=& {\Phi_0 \over 4\pi \lambda^2}
   [\, \ln \kappa + \alpha(\kappa) \,] \,,
        ~~ \nonumber
   h_{c1} = {B_{c1} \over B_{c2}} = { \ln \kappa +
   \alpha(\kappa) \over 2\kappa^2 }  \,,  \\
   \alpha(\kappa) &=& \alpha_\infty  + \exp[ - c_0
   - c_1 \ln\kappa - c_2 (\ln\kappa)^2] \pm \epsilon
   \end{eqnarray}
with $\alpha_\infty = 0.49693$, $c_0 = 0.41477$,
$c_1 = 0.775$, $c_2= 0.1303$, and $\epsilon \le 0.00076$.
This expression yields at $\kappa = 1/\sqrt 2$ the correct
value $h_{c1} = 1$ and for $\kappa \gg 1$
it has the limit $\alpha = 0.49693$.
A simpler expression for $\alpha(\kappa)$, yielding an
$h_{c1}$ with error still less than 1\% and with the correct
limits at $\kappa=1 / \sqrt2$ and $\kappa \gg 1$, is
  \begin{eqnarray}  
  \alpha(\kappa) =0.5 + (1+\ln 2 )/(2\kappa -\sqrt2 +2)\,.
  \end{eqnarray}
The resulting magnetization curves $M = \bar B -B_a$ are
shown in Fig.~3. They are well fitted by
  \begin{eqnarray}  
  h(b,\kappa) &=& {B_a \over B_{c2}} \approx h_{c1} +
   {c_1 b^3 \over 1 +c_2 b + c_3 b^2 }  \,, \nonumber \\
   c_1 &=& (1-h_{c1})^3 / (h_{c1}-p)^2  \,, \nonumber \\
   c_2 &=& (1-3h_{c1} +2p) / (h_{c1}-p) \,, \nonumber \\
   c_3 &=& 1 +(1-h_{c1})(1\!-2h_{c1}\!+p) /(h_{c1}\!-p)^2
  \end{eqnarray}
with $h_{c1}$ from Eq.~(56) and
$p=-{\rm d}m/{\rm d}b|_{b=1} =1/[(2\kappa^2\!-1)\beta_A +1]$
($m=b-h = M/B_{c2}$),
$\beta_A=1.15960$ (1.18034) for the triangular (square) vortex
lattice. This $h(b)$ satisfies the exact relations:
$h(0) =h_{c1}$, $h'(0)\!=h''(0)\!=h''(1)\!=0$,
$h(1) = 1$,  $h'(1) = 1-p(\kappa)$. The $M(B_a)$ from the
fit (58) applies for not too large $\kappa \le 10 \dots 20$.
For larger $\kappa$, better fits are given in Ref.~\cite{15},
where it is also shown that the often used ``logarithmic law
at $B_{c1} \ll B_a \ll B_{c2}$'' for the magnetization
$M(B_a) =\bar B -B_a$ has very limited range of validity.

  Figure 4 shows the profiles $B(r)$ and $\omega(r)=|\psi(r)|^2$
for an isolated vortex line $\bar B \to 0$ from GL theory for
$\kappa =$ 2, 5, and 20. Profiles for the vortex lattice
are shown in Fig.~5 for $\kappa = 5$ and for two inductions at
which the vortex spacing is $a=2\lambda$ and $a=4\lambda$.
Contour lines for $\omega(x,y)$ and $B(x,y)$ are plotted in
Fig.~6 for $\kappa =5$ at $b=0.5$. At $b > 0.7$ the contours
of $\omega(x,y)$ (see right hand part in Fig.~2) and $B(x,y)$
practically coincide, and at $b < 0.3$ the contours are nearly
circular, around well separated vortex cores and field peaks.

  The variance of the magnetic field
  \begin{eqnarray}  
  \sigma = \langle [ B(x,y) -\bar B]^2 \rangle =
  \sum_{{\bf K}\ne 0} B_{\bf K}^2
  \end{eqnarray}
is shown in Fig.~7. In the low-field range
$0.13/\kappa^2 \ll b \ll 1$ one has for the triangular lattice
the London limit $\sigma =0.00371 \Phi_0^2/\lambda^4 $
(upper frame in Fig.~7), at very small $b \ll 0.13/\kappa^2$
one has $\sigma =(b \kappa^2 /8\pi^2) \Phi_0^2/\lambda^4 $
(dash-dotted straight lines in in Fig.~7), and near $b=1$
one has the Abrikosov limit $\sigma =7.52\cdot 10^{-4}
(\Phi_0^2/\lambda^4) [\kappa^2 (1-b) / (\kappa^2 -0.069)]^2$.
This field variance is needed, e.g., for the interpretation
of Muon Spin Rotation ($\mu$SR) experiments \cite{19,20,21,22}.

   In Fig.~8 the shear modulus of the bulk triangular
vortex lattice is plotted versus the reduced induction for
various GL parameters $\kappa$. Note that for $\kappa =1/\sqrt2$
where $B_{c1}=B_c=B_{c2}=\Phi_0/(4\pi \lambda^2)$,
one has $c_{66} = 0$. One can
show that in the particular case $\kappa = 1/\sqrt2$ all
possible vortex configurations have the same free energy
$F= \bar B B_{c1}/\mu_0$, e.g., triangular and square lattice,
lattices with two flux quanta per vortex, or all vortices
merged into one giant vortex.

\section{Vortex lattice for thin and thick films} 

    The 2D Fourier method for the bulk vortex lattice in
Sec.~9 can be generalized to the 3D problem of vortex lattices
in infinite films of arbitrary thickness $d$ put into a
uniform magnetic field ${\bf B}_a$, since then the functions
$\omega(x,y,z)$ and ${\bf B}(x,y,z)$ are still periodic in
the $(x,y)$ plane. I consider here the case when
${\bf B}_a = B_a {\bf \hat z}$ is perpendicular to the film
plane \cite{23}, though in principle the Fourier method
applies also to tilted applied field. The total free energy
$F_{\rm tot}$ per unit volume of the infinite film is the
free energy, Eq.~(2), plus the stray-field energy
$F_{\rm stray}$, i.e., the energy of the magnetic field
variations outside the film,
  \begin{eqnarray}  
  F_{\rm tot} = F + {F_{\rm stray} \over d}\,,~~
  F_{\rm stray} = 2\int_{d/2}^\infty \! \langle {\bf B(r)}^2
    - \bar B^2 \rangle_{x,y} dz \,.
  \end{eqnarray}
One has $\bar B = B_a$ since all field lines have to cross
the infinite film. The factor of 2 in (60) comes from the
two half spaces above and below the film, which contribute
equally to $F_{\rm stray}$. The stray field
${\bf B}(x,y,z>d/2)$ with constant planar average
$\langle {\bf B}(x,y,z) \rangle_{x,y} = \bar B{\bf \hat z}$
is determined by the Laplace equation $\nabla^2 {\bf B} =0$
(since $\nabla\cdot {\bf B} =0$ and $\nabla \times {\bf B}=0$
in vacuum) and by its perpendicular component at the film
surface $z= d/2$, since $B_z$ has to be continuous across
this surface. The trial functions for $\omega({\bf r})$,
${\bf B(r)} = \bar B {\bf \hat z} + {\bf b(r)}$,
and ${\bf Q(r)} = {\bf Q}_A(x,y) + {\bf q(r)}$ inside the
film ($|z| \le d/2$) are now 3D Fourier series \cite{23},
 \begin{eqnarray}  
  \omega({\bf r}) &=& \sum_{\bf K} a_{\bf K} (1 - \cos
  {\bf K}_\perp {\bf r}_\perp ) \cos K_z z \,,
       \nonumber \\ \nonumber
  b_z({\bf r}) &=& \sum_{\bf K} b_{\bf K} \cos
  {\bf K}_\perp {\bf r}_\perp  \cos K_z z \,,
       \\
  {\bf b}_\perp({\bf r}) &=& \sum_{\bf K} b_{\bf K}
      {{\bf K}_\perp K_z \over K_\perp^2}
  \sin{\bf K}_\perp {\bf r}_\perp  \sin K_z z \,,
        \nonumber \\
  {\bf q}({\bf r}) &=& \sum_{\bf ~K} b_{\bf K}
  { {\bf \hat z \times K}_\perp \over K^2_\perp }
  \sin{\bf K}_\perp {\bf r}_\perp \cos K_z z \,,
  \end{eqnarray}
Here ${\bf r} = (x,y,z)$, ${\bf r}_\perp = (x,y)$,
${\bf K} = (K_x,K_y,K_z)$ with ${\bf K}_\perp = (K_x,K_y)$
from Eq.~(8), and $K_z = (2\pi/d)l$, $l=0,1,2, \dots$.
In all sums here and below the term ${\bf K}_\perp =0$ is
excluded. Minimizing $F_{\rm tot}$ with respect to the
coefficients $a_{\bf K}$ and $b_{\bf K}$ and using the
appropriate orthonormality relations one arrives at
iteration equations for the $a_{\bf K}$ and $b_{\bf K}$
similar to Eqs.~(52)-(54). The solution is then obtained
by first finding the 2D bulk solution as in Sec.~10
by considering only the terms with $K_z =0$. The magnetic
field lines then still have an unphysical sharp bend
at the surface, see left part of Fig.~9. Next we allow
for the terms with $K_z \ne 0$. This yields a
``mushrooming'' of the field lines of each vortex
when it approaches the film surface such that these
lines smoothly cross the surface with no bend, see
right part of Fig.~9.

  Profiles $\omega(x,0,z)$ and $B_z(x,0,z)$ are shown
in Fig.~10 for $z=0$ (middle plane of the film) and
$z=d/2$ (film surface). One can see that the spatial
variation of $B$ at the surface is reduced from its
bulk value by nearly $1/2$. Outside the film, the
transverse field components $B_x$, $B_y$ rapidly
decrease as $\exp(-|{\bf K}_{10}| z' )\approx
\exp(-2 \pi z' /a)$ where $z' = |z| -d/2$ is the
distance from the surface. Interestingly, the
profile of the order parameter $\omega$ in
films is almost independent of $z$.

  The shear modulus $c_{66}$ of the triangular vortex
lattice in films can be positive even when
$\kappa < 1/\sqrt2$, provided the film thickness $d$
is smaller than the coherence length $\xi$, see
Fig.~11. This means that a stable vortex lattice may
exist in thin type-I superconductor films. Our
numerical result confirms the $c_{66}$ of films that
was calculated analytically near $B_{c2}$ by
Albert Schmid \cite{24}.

  In such infinitely extended films one has $\bar B = B_a$
since all field lines have to pass the film. Therefore, the
magnetization $M$ of the film cannot be calculated as a
difference of fields, but one has to take the derivative
of the total free energy,
$M =\bar B -\partial F_{\rm tot} /\partial \bar B$.
A more elegant method calculates $M$ by Doria's virial
theorem \cite{18}, which for bulk superconductors yields
Eq.~(55). Indeed, it was shown recently \cite{25} that this
virial idea can be generalized to films of arbitrary thickness
and $M$ may be calculated directly from the GL solution for the
film, with no need to take an energy derivative. The resulting
magnetization of the film is plotted versus the applied field
in Fig.~12 for various film thicknesses $d$ and various GL
parameters $\kappa$. The curves for various $d$ cut each other
at $b\approx 0.1/\kappa$. For $\kappa = 0.707$ the thick film
limit ($d \gg \xi$ but still film width $w \gg d$) yields
a straight line, $-M/B_{c2} = 1-b$. This result is valid for
large demagnetization factor $N$, $1-N \ll 1$; it differs from
the bulk result for $M$ shown in Fig.~3, which is valid for
the demagnetization-free limit of $N \ll 1$.

\section{Remarks on the magnetization} 

  One may ask why the magnetization ${\bf M = m}/V$
is not calculated via the general definition of the
magnetic moment ${\bf m} = V{\bf M} = {1\over2} \int_V
 d^3r\, {\bf r \times j}\,$ \cite{26,27} from the current
density ${\bf j(r)}$, which is easily calculated as a periodic
function by our Fourier method, both for bulk and film
superconductors. However, into this definition enters not only
the periodic part, i.e., the vortex currents circulating inside
each vortex cell; this contribution even would give the wrong
sign of ${\bf m}$. The main contribution to the magnetic moment
of a superconductor of any shape comes from the screening
currents that flow near the surface of the specimen.
The magnetization in superconductors is thus not a volume
property as it is in magnets. For example,
for a long cylinder in parallel field $B_a$ the magnetization
$M = \bar B - B_a < 0$ is composed of the positive contribution
$\bar B$ of the vortex currents and the negative (and larger)
contribution $B_a$ of the surface currents that screen the
cylinder from the applied field $B_a$ before vortices are
allowed to penetrate. Near $B_{c2}$, both terms nearly
compensate and $|M|$ is a small difference of two big terms.
In the film geometry, and for most other shapes of the
superconductor, the screening current is not easily known
but has to be computed; such computations for thin and thick
strips, disks, and plates of macroscopic size $\gg \xi$ are
presented in \cite{26}. Landau knew this problem, since he
worked on demagnetization factors and on the intermediate
state in type-I superconductors \cite{27}, and he used the
thermodynamic definition of the magnetization as an energy
derivative. In my view, the simple formula (55) derived by
Doria, Gubernatis and Rainer \cite{18} by scaling the
coordinates in GL theory and finding a novel virial
relationship between kinetic and potential energies from
which the equilibrium field $B_a = \bar B -M$ follows,
was a fundamental discovery that occurred long after the
publication of GL theory \cite{1}.

\end{document}